\documentclass[%
aip,
amsmath,amssymb,
reprint,%
]{revtex4-1}

\usepackage{graphicx}
\usepackage{dcolumn}
\usepackage{bm}
\usepackage{hyperref}
\usepackage[dvipsnames]{xcolor}
\usepackage[utf8]{inputenc}
\usepackage[T1]{fontenc}
\usepackage{mathptmx,mathtools}

\begin{document}
\preprint{AIP/123-QED}

\title[]{Dynamics of coupled modified R\"ossler oscillators: the role of nonisochronicity parameter}

\author{C. Ramya}
\author{R. Gopal}
\author{R. Suresh}
\email{sureshphy.logo@gmail.com.}
\author{V. K. Chandrasekar}%
\email{chandru25nld@gmail.com.}
\affiliation{ 
	Department of Physics, Centre for Nonlinear Science and Engineering, School of Electrical and Electronics Engineering, SASTRA Deemed University, Thanjavur  613 401, India
}%
\date{\today}

\begin{abstract}{The amplitude-dependent frequency of the oscillations, termed \emph{nonisochronicity}, is one of the essential characteristics of nonlinear oscillators.  In this paper, the dynamics of the R\"ossler oscillator in the presence of nonisochronicity is examined. In particular, we explore the appearance of a new fixed point and the emergence of a coexisting limit-cycle and quasiperiodic attractors. We also describe the sequence of bifurcations leading to synchronized, desynchronized attractors and oscillation death states in the coupled R\"ossler oscillators as a function of the strength of nonisochronicity and coupling parameters.  Further, we characterize the multistability of the coexisting attractors by plotting the basins of attraction.  Our results open up the possibilities of understanding the emergence of coexisting attractors, and into a qualitative change of the collective states in coupled nonlinear oscillators in the presence of nonisochronicity.}
\end{abstract}
\maketitle
\begin{quotation}
Nonisochronicity is one of the essential and peculiar characteristic property of the nonlinear differential equations. We show here that for R\"ossler oscillator with nonisochronicity term, there exist many distinct dynamical states, due to the appearance of a new fixed point, which leads to rich and complex dynamics in the system. We observe that the coexistence of many attractors may introduce additional sensitivity and randomness in the chaotic system because an infinitesimal perturbation in initial values is sufficient for a trajectory to shift from one state to another. This can be of great interest for secure communication applications through encryption of information-based chaos, for example, and various control strategies in engineering contexts. We describe and characterize the regions of synchronization in coupled R\"ossler oscillators,  basins of attraction corresponding to different attractors using master stability function formalism. We also observe that the system's permutational symmetry is broken at a critical value of the coupling strength, which persuades multi-stable synchronized, desynchronized attractors, and oscillation death states in the coupled Rössler oscillators.
\end{quotation}
\section{Introduction}
\label{sec1}
Systems that are described by the nonlinear differential equations are omnipresent ranging from biological systems to physical systems and economics\cite{strogatz2018,leung1989,puu2003}. The nonlinearities present in these systems facilitate the emergence of rich dynamical states such as self-sustained oscillations, turbulence, and chaos, which are impossible to obtain in the systems described by the linear differential equations. In particular, one of the important characteristics of nonlinear differential equations is the amplitude-dependent frequency of nonlinear oscillations, known as nonisochronicity, the term first used in ref.\cite{aronson1990}. Nonisochronicity is central to inhomogeneities that can be seen in a variety of living situations \cite{daido2006,tiberkevich2014} and due to this nonisochronous nature of the nonlinear equations, different responses have been observed for different input conditions in practical or experimental settings. In particular, when considering the complex biological or other networks, the strong nonisochronicity often results in rich synchronization patterns and the level of isochronicity of oscillators is an important property of the phase interaction function, where the strong dependence of the oscillatory period or frequency on the amplitude results in high-level shear in the phase dynamics and coupling can dramatically change the oscillatory period\cite{sebek2016}. Also, it has been reported that the higher nonisochronicity level can induce long, irregular transient phase dynamics in the networks, where such long transient dynamics have relevance in the functioning of biological systems\cite{wickramasinghe2013}. The detection of isochrons in a system is also prescribed recently\cite{mauroy2012,mauroy2013}. Moreover, nonisochronicity plays a significant role in the development of anomalous phase synchronization in populations of nonidentical oscillators\cite{blasius2003}, which has also been experimentally confirmed\cite{dana2006,wickramasinghe2011}. Further, in the networks of nonlocally coupled Stuart-Landau oscillators, the nonisochronicity leads to the emergence of various collective states such as amplitude chimera, amplitude cluster, frequency chimera, frequency cluster states, imperfect and mixed imperfect synchronized states, local and global chimera states, etc., which have been recently reported in the case of symmetry preserving and symmetry breaking couplings \cite{premalatha2015,premalatha2016a,premalatha2016b,premalatha2018,senthilkumar2019}. Similarly, the emergence of nonisochronicity induced chimera, and chimera-like states in nonlocally and globally coupled complex Ginzburg-Landau model\cite{kuramoto2002,sethia2014,mishra2015}, van der Pol
system\cite{hens2015} are also demonstrated. Furthermore, in ref.\cite{montbrio2003}, the authors reported that the nonisochronicity could also be used to control the transition to synchronization in an ensemble of nonidentical oscillators.

As mentioned earlier, the strong nonisochronicity parameter induces rich collective dynamics in a coupled system of oscillators. However, the role of nonisochronicity has been mainly studied in the case of Stuart-Landau oscillator, coupled complex Ginzburg-Landau model and only a few studies are reported in other dynamical systems. Nevertheless, there are many other interesting nonlinear oscillatory systems and biological models that deserve to be studied in detail in the presence of nonisochronicity. In compliance with this, in a recent study \cite{chandrasekar2014}, the authors have reported the advent of chimera states due to the presence of nonisochronicity term in a diffusively coupled global network of R\"ossler oscillators. They also found that the presence of multistability\cite{sommerer1993,feudel1998,do2008,lai2017,wontchui2017,hens2012,jaros2015,patel2014} is the key mechanism for the appearance of coexisting states. Moreover, the emergence of chimera state in a star network of R\"ossler oscillators with nonisochronicity term is demonstrated with various coupling topologies\cite{meena2016}.

Further, in Ref. \cite{chaurasia2017}, the authors have explored the possibility of controlling chaos in an ensemble of modified R\"ossler oscillators, in which the individual nodes are indirectly coupled through a single external oscillator. Furthermore, the investigation of the emergence of chaotic phase synchronization in three coupled modified R\"ossler oscillator systems with parameter mismatches are reported\cite{nishikawa2009}. The authors showed the switching between complete and partial phase synchronization as a function of the coupling strength. The transition from in-phase to out-of-phase synchronizations via the phase-flip transition in a network of relay-coupled modified R\"ossler oscillators is demonstrated by Sharma et al.\cite{sharma2011}. In all the studies mentioned above, the R\"ossler oscillator with nonisochronicity term has been used to demonstrate the results. However, the basic knowledge of the dynamical changes that occurred in the R\"ossler oscillator with nonisochronicity parameter is still lagging. Additionally, the study of nonisochronicity term induced bifurcations and collective dynamical states exhibited in the coupled systems with different coupling topologies are required exhaustive investigation. 

Taking this as our motivation, in this paper, we study the dynamics of the R\"ossler oscillator by adding the nonisochronicity term. Without which the system exhibits limit-cycle oscillations for the chosen parameter values. When we incorporate the nonisochronicity term into the oscillator equation, the system manifests the coexistence of limit-cycle and quasiperiodic attractors. This is due to the appearance of a new fixed point in response to the strength of the nonisochronicity parameter. Next, the study is extended to a system of two mutually coupled R\"ossler oscillators, and we investigate the influence of the coupling parameter along with the nonisochronicity term. We show that when one increases the strength of coupling, by fixing the nonisochronicity parameter, the permutational symmetry (($x_1, y_1, z_1$)$\longleftrightarrow$($x_2, y_2, z_2$)) of the subsystems is broken at a critical value of the coupling strength, that persuades multistable desynchronized attractors, which are emerged via different bifurcations based on the strength of nonisochronicity and coupling parameters. 

When the coupling strength reaches a sufficiently larger value, the symmetry is preserved again, which leads one of the attractors to exhibit synchronized oscillations. Here, the synchronization is defined as the correlation of both phase and amplitude of the oscillations. During the occurrence of desynchronized oscillations, either amplitude or both phase and amplitude are uncorrelated. In addition to the synchronized and desynchronized oscillations, we discover the development of two distinct oscillation death (OD) states (in which the subsystems exhibit inhomogeneous steady-state solutions) along with the synchronized and desynchronized oscillations. The occurrence of steady state solutions \cite{prasad2005,saxena2012,resmi2011,resmi2012,	  hens2013,nandan2014,verma2018,verma2019} and multistability \cite{hens2012,jaros2015,patel2014} has been reported in networks of original R\"ossler oscillators with different coupling configurations. However, we emphasize that to the best of our knowledge, the emergence of OD states and coexisting attractors are unexplored yet in the coupled R\"ossler oscillators with nonisochronicity term. Further, the multistability is characterized by estimating the bifurcation diagrams, and the region of complete synchronization is identified using the master stability function (MSF) formalism. Also, based on the emergence of different attractors, the bifurcation plot is divided into several regions. The development of coexisting attractors in those regions is corroborated by plotting the basins of attraction.

The structure of the remaining paper is organized as follows: In Sec.~\ref{sec2}, we will introduce the mathematical model of the R\"ossler oscillator with nonisochronicity term, and discuss the stability of the fixed points of the system with and without the nonisochronicity term. In Sec. \ref{sec3}, we study the dynamics of a R\"ossler oscillator with nonisochronicity term and show the emergence of coexisting attractors that are emerged via various bifurcations. Next, in Sec. \ref{sec4}, we will consider two mutually coupled oscillators for our study and show the emergence of coexisting synchronized, desynchronized attractors along with OD states using bifurcation diagrams. Further, the bifurcation diagram is divided into different regions based on the combination of coexisting attractors, and the multistability is confirmed by plotting the basins of attraction in Sec.~\ref{sec5}. Moreover, the probability of occurrence of each attractor in the basins is also discussed. Finally, the obtained results are consolidated, and the future directions of research are discussed in Sec.~\ref{sec7}.
\section{Mathematical model and fixed point analysis}
\label{sec2}
In order to demonstrate our results, we first consider a R\"ossler oscillator with nonisochronicity term, which can be described by the mathematical equation of the form 
\begin{eqnarray}
\label{eqn1}
\dot{x}&=& -\omega_0(1-\alpha (x^2+y^2))y-z, \nonumber\\
\dot{y}&=& \omega_0(1-\alpha (x^2+y^2))x+ay,\\
\dot{z}&=& b+(x-c)z \nonumber,
\end{eqnarray}
where $x,y$, and $z$ are the state variables, $\omega_0$ is the natural frequency of the system, $a,b$, and $c$ are the control parameters. $\alpha$ is a parameter to adjust the strength of nonisochronicity.

\begin{figure}
	\includegraphics[width=1.0\columnwidth]{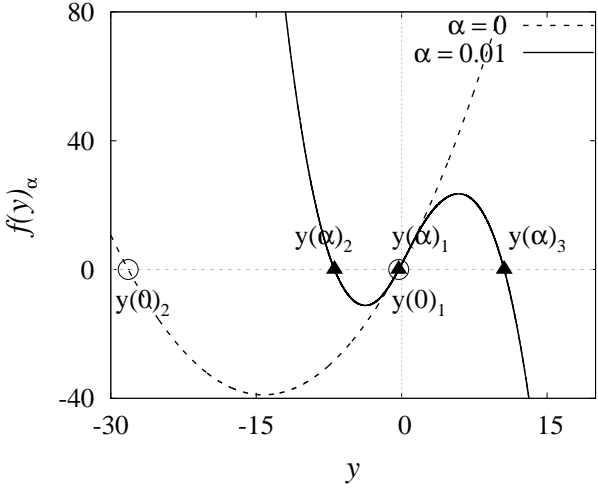}
	\caption{Fixed points of the $y$--component of a R\"ossler oscillator (\ref{eqn1}) in the absence of nonisochronicity parameter ($\alpha=0$) and for $\alpha=0.01$. The system parameters are fixed at $\omega_0=1$,  $a=0.2$, $b=1.7$ and $c=5.7$.}
	\label{rossler_fp}
\end{figure}

In order to find the fixed points, the three R\"ossler equations are set to zero and the resulting equations of $x$ and $z$--components can be written in terms of the $y$--variable as follows,
\begin{eqnarray}
\label{eqn2}
x&=&\frac{acy^{2}}{ay^2+b}, \quad z=\frac{ay^2+b}{c},\\
f(y)_\alpha&=&cy\omega_0\left(1-\alpha\left(\frac{a^2c^2y^4}{\left(ay^2+b\right)^2}+y^2\right)\right)+ay^2+b=0\nonumber.
\end{eqnarray}
For $\alpha=0$, the function $f(y)_\alpha$ in Eq.~(\ref{eqn2}) becomes 
\begin{equation}
\label{eqn2a}
f(y)_0=y(ay+c\omega_0)+b,
\end{equation}
and the corresponding two roots (fixed points) are given by
\begin{equation}
\label{eqn2b}
y(0)_{1,2}=\pm\frac{\sqrt{c^2\omega_{0}^2-4ab}}{2a}-c\omega_{0}.
\end{equation} 
The respective two fixed points $y(0)_1$ and $y(0)_2$ are depicted as open circles in Fig.~\ref{rossler_fp} for the fixed values of the control parameters $\omega_0=1, a=0.2, b=1.7$ and $c=5.7$. When $\alpha\neq0$, the function $f(y)_\alpha$ has three roots (fixed points) $y(\alpha)_{1,2,3}$. The fixed points $y(\alpha)_{1,2}$ are the same fixed points that were present when $\alpha=0$, however the positions are now dependent on the value of $\alpha$ (see filled triangles in Fig.~\ref{rossler_fp}). They are approximately given by the following form
%
\begin{eqnarray}
y(\alpha)_{(1,2)}\approx\bigg(a^2\beta^5\left(5a\beta-2c\omega_0\left(2\alpha c^2+3\alpha\beta^2-2\right)\right)+\nonumber\\b^2\beta^2(3a-2\alpha c\omega_0\beta)+ab\beta^3\left(9a\beta+4c\omega_0\left(1-2\alpha\beta^2\right)\right)-b^3\bigg)\nonumber \\ \times\bigg(6a^3\beta^5+a^2\beta^3\left(12b+c\omega_0\left(-5\alpha c^2-7\alpha  \beta^2+5\right)\right)\nonumber \\+2ab\beta\bigg(3b+c\omega_0\beta\left(3-5\alpha\beta^2\right)\bigg)+b^2c\omega_0\left(1-3\alpha\beta^2\right)\bigg)^{-1}
\label{model1}
\end{eqnarray}
where $\beta=y(0)_{(1,2)}$. To find the fixed points of Eq. (\ref{eqn1}), the nonisochronicity parameter $\alpha$ is considered as a small perturbation. The other fixed point $y(\alpha)_{3}$ is newly emerged upon introducing $\alpha$. This can be clearly seen from Fig.~\ref{rossler_fp} that the new fixed point has emerged in the positive $y$-direction due to the effect of $\alpha$. For the aforementioned values of control parameters, the three fixed points are unstable and the nature of their eigenvalues are 
\begin{enumerate}
	\item FP$(\alpha)_{1}=(x(\alpha)_1, y(\alpha)_1, z(\alpha)_1)$ -- which has one negative real eigenvalue (near zero) and one complex conjugate eigenvalue pair with a positive real part.
	\item FP$(\alpha)_{2}=(x(\alpha)_2, y(\alpha)_2, z(\alpha)_2)$ -- which has one complex conjugate eigenvalue pair with a negative real part and one positive real eigenvalue.
	\item FP$(\alpha)_{3}=(x(\alpha)_3, y(\alpha)_3, z(\alpha)_3)$ -- which has one negative real eigenvalue and one complex conjugate eigenvalue pair with a positive real part. 
\end{enumerate}
\begin{figure}
	\includegraphics[width=1.0\columnwidth]{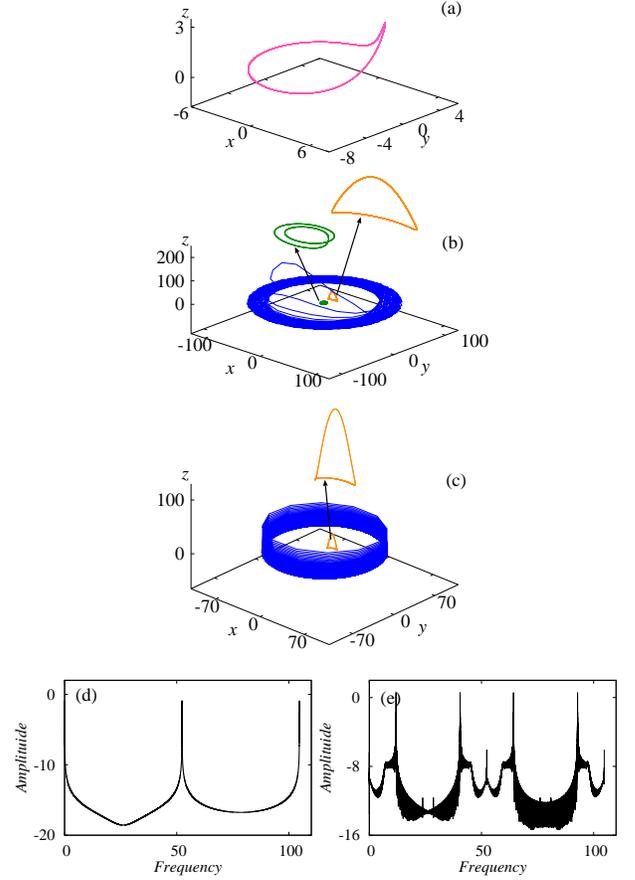}
	\caption{The attractors that are emerged for different values of the nonisochronicity parameter ($\alpha$) in the R\"ossler oscillator (\ref{eqn1}). Other parameters are fixed as shown in Fig.~\ref{rossler_fp}. (a) Three dimensional projection of the period-I limit-cycle attractor emerged for $\alpha=0$, (b) shows the period-II attractor along with the coexisting period-I attractor and large-amplitude quasiperiodic attractor for different sets of initial states, for $\alpha=0.007$, (c) depicts the coexistence of period-I limit-cycle and quasiperiodic attractors that are appeared in the system for $\alpha=0.01$. (d) and (e) Depicts the power spectrum of the periodic and quasiperiodic attractors (shown in Fig.\ref{single_ros_att}(c)), respectively.}
	\label{single_ros_att}
\end{figure} 
When $\alpha=0$, the system shows a single limit-cycle attractor, for the above chosen parameter values, which is rotating about the fixed point $y(0)_{1}$. When introducing $\alpha$, due to the emergent of a new fixed point $y(\alpha)_{3}$, the system shows additional limit-cycle and quasiperiodic attractors. We explain this in detail in Sec. \ref{sec3}. One can also notice that in the case of $c^{2}\omega_{0}^{2}<4ab$, the system (\ref{eqn1}) has zero (one) fixed point for $\alpha=0$ ($\neq 0$). 
\begin{figure}
	\includegraphics[width=1.0\columnwidth]{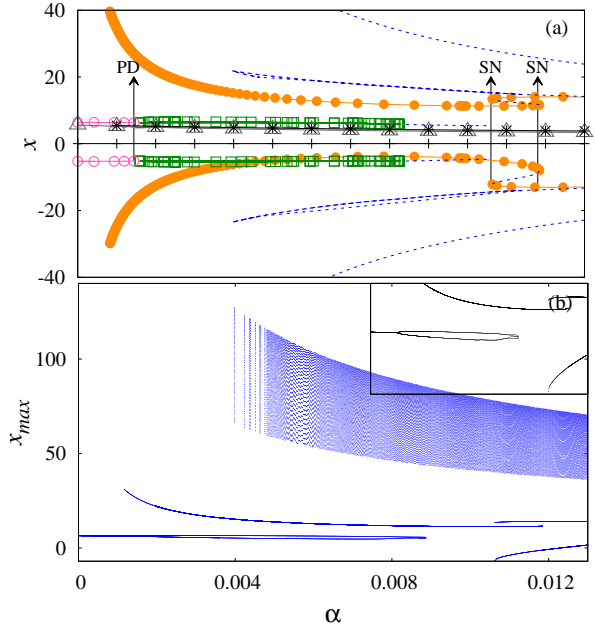}
	\caption{(a), (b) Bifurcation diagrams of a single R\"ossler oscillator model (\ref{eqn1}) as a function of the nonisochronicity parameter $\alpha\in[0,0.013]$, shows the emergence of different dynamical states. The evolution of three fixed ponts are depicted as lines with open triangle, asterisk, and plus symbol. Other system parameters are fixed as given in Fig.~\ref{rossler_fp}.}
	\label{single_ros_bif}
\end{figure} 
\section{Dynamics of an individual R\"ossler oscillator with nonisochronicity}
\label{sec3}
First, we study the dynamics of Eq.~(\ref{eqn1}) as a function of the nonisochronicity parameter ($\alpha$). For our numerical investigation, we have integrated Eq.~(\ref{eqn1}) using the fourth-order Runge-Kutta method with a step size of 0.01. We also confirm that the similar results can be observed when we use smaller time-steps of 0.001. The emerging attractors are depicted in Fig.~\ref{single_ros_att} for three selective values of $\alpha$. First, in the absence of $\alpha$, as mentioned earlier, the R\"ossler oscillator has two fixed points (FP(0)$_{1}$ and FP(0)$_{2}$), in that FP(0)$_{2}$ is saddle-focus whose one-dimensional unstable manifold repels the trajectories along the stable manifold of the other fixed point FP(0)$_{1}$, which is also a saddle-focus. Therefore, a period-I limit-cycle attractor is emerged, which rotates (counter-clockwise) around FP(0)$_{1}$ fixed point. The three-dimensional projection of the respective limit-cycle attractor in the $(x,y,z)$ phase plane is displayed in Fig.~\ref{single_ros_att}(a) for $\alpha=0$. When we incorporate and slowly increase the nonisochronicity parameter further, the period-I limit-cycle is then bifurcated into a period-II attractor via period-doubling bifurcation. At the same time, due to the inclusion of $\alpha$, one more new fixed point FP($\alpha$)$_{3}$ has emerged in the system, as shown in Fig.~\ref{rossler_fp}. Consequently, another period-I limit-cycle appears for different sets of initial states, which rotate around the fixed point FP($ \alpha $)$_{3}$. In addition to this, a large-amplitude quasiperiodic attractor is also appeared and rotating around the fixed points FP($\alpha$)$_{1}$, FP($\alpha$)$_{2}$ and FP($\alpha$)$_{3}$. All three coexisting attractors that are emerged for different sets of initial states are depicted in Fig.~\ref{single_ros_att}(b) for the value of $\alpha=0.007$. The periodic (period-I attractor) and quasiperiodic nature of the attractors are confirmed by estimating the power spectrum and are plotted in Figs.\ref{single_ros_att}(d), and \ref{single_ros_att}(e), respectively. The period-II attractor has become unstable when $\alpha$ reaches a threshold value. Therefore, only the other two attractors are stable and coexisted in the system, which is depicted in Fig.~\ref{single_ros_att}(c) for $\alpha=0.01$. 

Overall, the R\"ossler oscillator (\ref{eqn1}) exhibits coexisting limit-cycle and quasiperiodic attractors when we incorporate the nonisochronicity parameter in it. To acquire a clear picture of the dynamics of the system, the bifurcation diagram of Eq~(\ref{eqn1}) is depicted in Fig.~\ref{single_ros_bif}(a) as a function of $\alpha\in[0,0.013]$, which is obtained using the XPPAUT software package\cite{ermentrout2002}. In this figure, the lines with different points represent various stable attractors, and the broken lines indicate the unstable periodic orbits.  In particular, the lines with open circles in Fig.~\ref{single_ros_bif}(a) indicate the maxima and minima of the period-I limit-cycle oscillations equivalent to the attractor shown in Fig.\ref{single_ros_att}(a). The period-doubling bifurcation, which occurred at the critical value of $\alpha=0.0016$, is marked as PD in Fig.~\ref{single_ros_bif}(a). The lines with open squares indicate the maxima and minima of the respective period-II attractor, and this period-II orbit is then become unstable when $\alpha>0.0082$. The newly emerged period-I limit-cycle attractor for other sets of initial conditions is coexisted along with the period-II orbit is also depicted in Fig.~\ref{single_ros_bif}(a) as lines with solid circles. The labels SN indicates the saddle-node bifurcation. The evolution of the three fixed points (lines with open triangle, asterisk, and plus symbols) with respect to $\alpha$ is also depicted in Fig.~\ref{single_ros_bif}(a). Among them the newly emerged fixed point is depicted with asterisk symbol.

The emergence of multistable attractors for different set of initial conditions as a function of the nonisochronicity parameter can be confirmed again in Fig.~\ref{single_ros_bif}(b), in which the maxima of the system variable $x$ ($x_{max}$) are plotted as a function of $\alpha$, shows the qualitative changes that occurred in the system. The bifurcation of the perioic orbits and the emergence of quasiperiodic states can be clealry visualized in Fig.~\ref{single_ros_bif}(b) with respect to $\alpha$. The inset in Fig.~\ref{single_ros_bif}(b) elucidate the period-doubling phenomenon.
\begin{figure}
	\includegraphics[width=1.0\columnwidth]{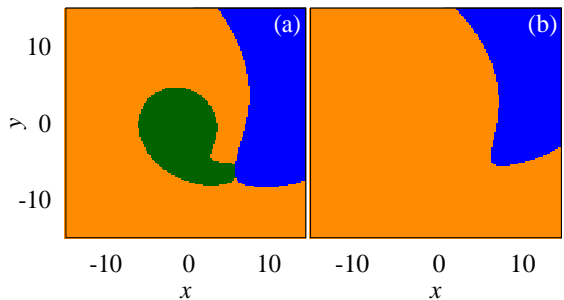}
	\caption{Basins of attraction of the system (\ref{eqn1}) as a function of the system variables $x$ and $y$ for (a) $\alpha=0.007$, and (b) $\alpha=0.01$ shows the emergence of coexisting attractors for different initial states.}
	\label{single_ross_basin}
\end{figure} 

Further, to identify the initial states to which different attractors emerge in the system, the basins of attraction of the R\"ossler oscillator is plotted in Figs.~\ref{single_ross_basin}(a) and \ref{single_ross_basin}(b) by fixing $z=-1$ for $\alpha$ = 0.007 and 0.01, respectively (corresponding to the attractors depicted in Figs.~\ref{single_ros_att}(b) and \ref{single_ros_att}(c)). In Fig.~\ref{single_ross_basin}(a), the green (dark gray) colored region indicates the basin of attraction of the period-II limit-cycle attractor, the area of orange (light gray) color represents the basin of attraction of the period-I limit-cycle attractor, and finally, the blue (black) color domain portrays the initial states to which the large-amplitude quasiperiodic attractor emerge in the system. We wish to emphasize here that, for our upcoming study, we choose the initial states only from the limit-cycle region and omit the initial states, which leads to the large-amplitude quasiperiodic oscillations.  

Thus, an individual R\"ossler oscillator exhibits coexisting attractors with respect to the strength of the nonisochronicity parameter. It is also of interest to investigate the dynamics of the coupled system and how the coupling parameter, along with the nonisochronicity term, influence the system dynamics. Therefore, in Sec.~\ref{sec4}, we will consider two mutually coupled system of R\"ossler oscillators and study the dynamics as a function of the coupling and nonisochronicity parameters.
%
\begin{figure*}
	\includegraphics[width=2.0\columnwidth]{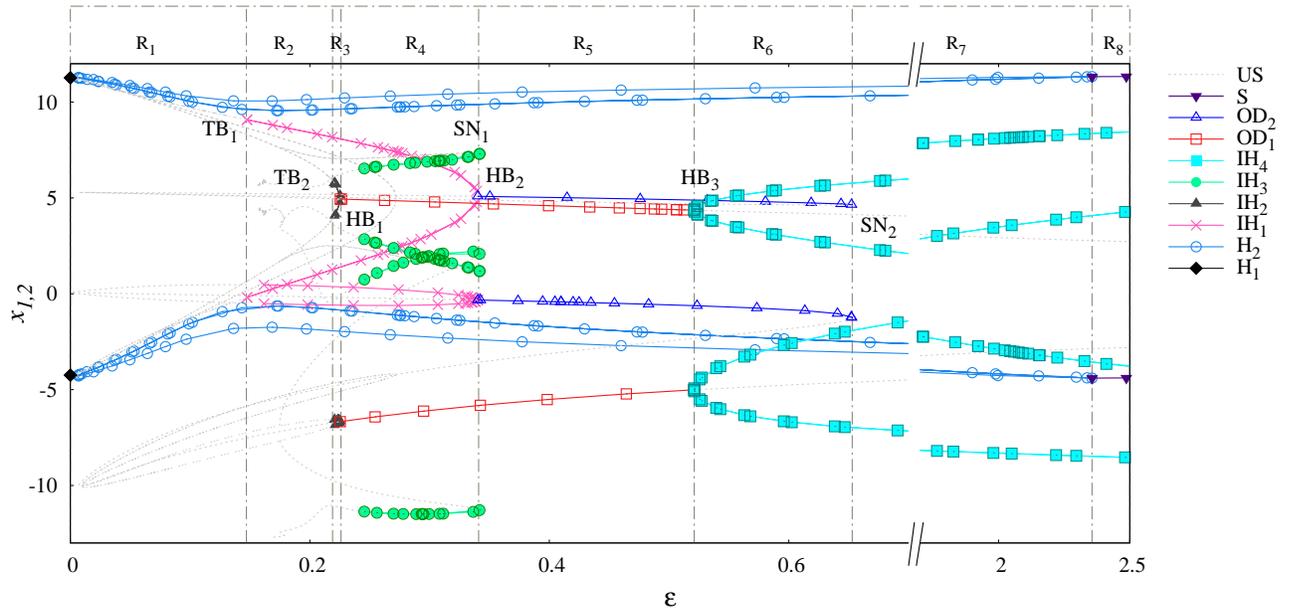}
	\caption{Bifurcation diagram of the system (\ref{eqn3}) illustrates the occurrence of different bifurcations and multiple attractors as a function of the coupling strength $\varepsilon\in[0,2.5]$ for $\alpha=0.01$. The vertical dashed lines indicate the critical values of $\varepsilon$, in which different bifurcations occur and separates the region from R$_1$ to R$_8$. Other parameters are fixed as given in Fig.~\ref{rossler_fp}.}
	\label{bifur_al0.01}
\end{figure*} 
\section{Dynamics of the coupled R\"ossler oscillator and the influence of nonisochronicity and coupling parameters}
\label{sec4}
The mathematical model of a system of two mutually coupled R\"ossler oscillators can be given as 
\begin{eqnarray}
\label{eqn3}
\dot{x}_{1,2}&=& -\omega_0(1-\alpha (x^2+y^2))y_{1,2}-
z_{1,2}+\varepsilon(x_{2,1}-x_{1,2}), \nonumber\\
\dot{y}_{1,2}&=& \omega_0(1-\alpha (x^2+y^2))x_{1,2}+ay_{1,2},\\
\dot{z}_{1,2}&=& b+(x_{1,2}-c)z_{1,2}\nonumber,
\end{eqnarray}
where $\varepsilon~(\geq0)$ represents the strength of diffusive coupling. In Eq.~(\ref{eqn3}), if we vary the coupling strength, by fixing $\alpha$ as a constant, the coupled system undergoes critical bifurcations. Consequently, different desynchronized attractors have emerged due to the breaking of the system's permutational symmetry and coexisted for different sets of initial states, including the birth of two different OD states. The one-parameter bifurcation diagram is estimated as a function of $\varepsilon$ by fixing $\alpha$ to display the development of different bifurcations and show the transition to complete synchronization state of the coupled system. 
\begin{table*}
\begin{center}
\caption{Table shows the emergence of different attractors in system(\ref{eqn3}) as a function of coupling strength $\varepsilon\in[0,2.5]$ for	nonisochronicity parameter $\alpha=0.01$.}
\label{tab1}       
\begin{tabular}{lcr}
\hline\noalign{\smallskip}
Attractor name & Oscillation type & Range of $\varepsilon\in[0,2.5]$ \\
\noalign{\smallskip}\hline\noalign{\smallskip}
H$_1$ &  Homogeneous oscillations--1 & 0 $<\varepsilon\leq$ 0.001\\
QP$_1$ & Quasiperiodic oscillations--1 & 0.001 $<\varepsilon\leq$ 0.147 \\
QP$_2$ & Quasiperiodic oscillations--2 & 0.001 $<\varepsilon\leq$ 0.22 \\
H$_2$ & Homogeneous oscillations--2 & 0.001 $<\varepsilon\leq$ 2.356\\
IH$_1$ & Inhomogeneous oscillations--1 & 0.147 $<\varepsilon\leq$ 0.34\\
IH$_2$ & Inhomogeneous oscillations--2 & 0.22 $<\varepsilon\leq$ 0.225\\
OD$_1$ & Oscillation death state--1 & 0.225 $<\varepsilon\leq$ 0.521\\
IH$_3$ & Inhomogeneous oscillations--3 & 0.245 $<\varepsilon\leq$ 0.341\\
OD$_2$ & Oscillation death state--2 & 0.341 $<\varepsilon\leq$ 0.653\\
IH$_4$ & Inhomogeneous oscillations--4 & $\varepsilon>$ 0.521\\
S & Synchronized oscillations & $\varepsilon>$ 2.356\\
\noalign{\smallskip}\hline
\end{tabular}
\end{center}
\end{table*}
%
\begin{figure}
	\includegraphics[width=1.0\columnwidth]{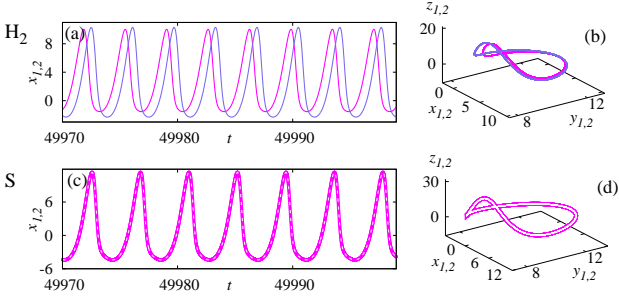}
	\caption{Time evolution and the corresponding phase portraits of the system (\ref{eqn3}) obtained for different initial states for $\alpha=0.01$. [(a), (b)] Homogeneous oscillations--2 (H$_2$) for $\varepsilon=0.14$,  and [(c), (d)] shows the synchronized oscillations (S) of the subsystems for $\varepsilon=3.0$.}
	\label{ts_att1}
\end{figure} 
\begin{figure}
	\includegraphics[width=1.0\columnwidth]{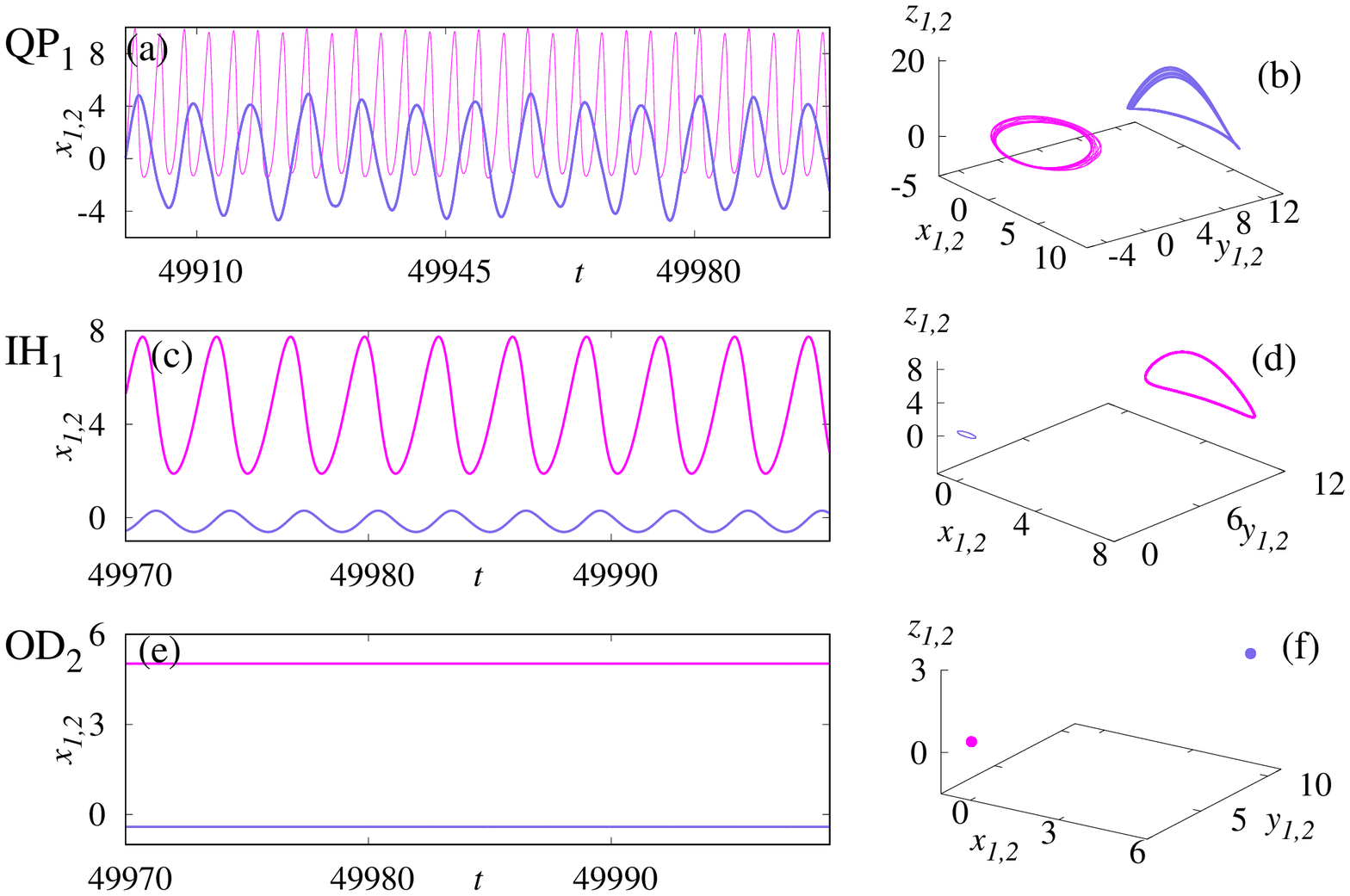}
	\caption{Time evolution and the respective phase portraits of the system (\ref{eqn3}) obtained for $\alpha=0.01$. [(a), (b)] Quasiperiodic oscillations--1 (QP$_1$) obtained for another set of initial states for $\varepsilon=0.14$, [(c), (d)] inhomogeneous oscillations--1 (IH$_1$) for $\varepsilon=0.25$, and [(e), (f)] shows the OD state--2 (OD$_2$) for $\varepsilon=0.4$.}
	\label{ts_att2}
\end{figure} 

The bifurcation diagram of the coupled system (\ref{eqn3}) is depicted in Fig.~\ref{bifur_al0.01} for $\varepsilon\in[0,2.5]$, and $\alpha=0.01$. Again, the stable orbits are identified using the lines with different points, and broken lines indicate the unstable orbits. The vertical dashed lines represent the critical values of $\varepsilon$, at which different bifurcations occur. Also, they separate the bifurcation diagram into different regions based on the bifurcations that occur. In the absence of coupling, both subsystems exhibit the coexistence of period-I limit-cycle and large-amplitude quasiperiodic attractors, as depicted in Fig.~\ref{single_ros_att}(c). In particular, the period-I limit-cycles of the subsystems oscillate with equal amplitude, satisfying the permutational symmetry ($x_1, y_1, z_1$)$\longleftrightarrow$($x_2, y_2, z_2$), but the phases are uncorrelated (exhibit constant phase shift), characterized as homogeneous oscillations--1 and the attractors are named as H$_1$. The solid diamonds in Fig.~\ref{bifur_al0.01} for $\varepsilon=0$ represent the maxima and minima of the homogeneous oscillations. When we slowly increase the coupling strength beyond $\varepsilon>0.001$, three different attractors have newly emerged in the system and coexisted for different initial states. These three attractors are further brought to the number of attractors as a function of the coupling strength, via different bifurcation routes. To be more specific, the systems exhibit desynchronized period-I limit-cycle attractor (H$_2$) along with two different quasiperiodic attractors, namely QP$_1$ and QP$_2$. The quasiperiodic dynamics is confirmed again by estimating the power spectrum. In the following, how these three attractors bifurcated and give birth to new attractors via different bifurcations as a function of $\varepsilon$ is explained.
\begin{figure}
	\includegraphics[width=1.0\columnwidth]{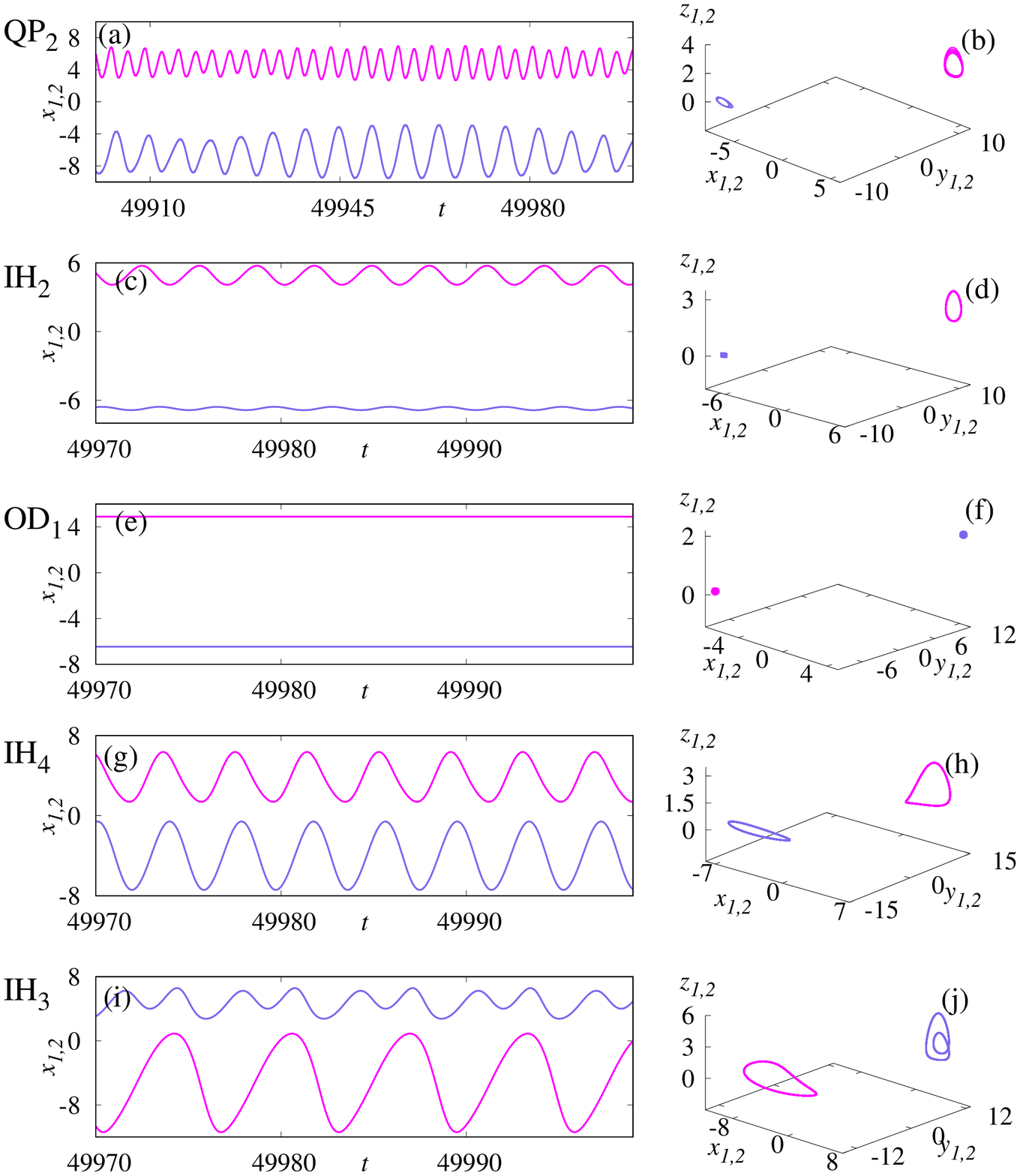}
	\caption{Time evolution and the equivalent phase portraits of the system (\ref{eqn3}) obtained for different initial states for $\alpha=0.01$. [(a), (b)] Quasiperiodic oscillations--2 (QP$_2$) for $\varepsilon=0.14$, [(c), (d)]  inhomogeneous oscillations--2 (IH$_2$) for $\varepsilon=0.221$, [(e), (f)] OD state--1 (OD$_1$) obtained for an another set of initial states for $\varepsilon=0.25$, [(g), (h)] inhomogeneous oscillations--4 (IH$_4$) for $\varepsilon=0.8$, and [(i), (j)] shows the emergence of inhomogeneous oscillations--3 (IH$_3$) for $\varepsilon=0.25$.}
	\label{ts_att3}
\end{figure} 
\begin{enumerate}
	\item In the attractor H$_2$, both the subsystems begin to evolve with different amplitudes (due to the breaking of permutational symmetry), but still rotating with the common center of rotation, which is called as homogeneous oscillations--2. The line with open circles in Fig.~\ref{bifur_al0.01} indicates the maxima and minima of the attractor H$_2$. The time series and related phase space plots of the desynchronized attractor H$_2$ are depicted in Figs. \ref{ts_att1}(c) and \ref{ts_att1}(d), respectively, for $\varepsilon=0.14$. The attractor H$_2$ is stable until $\varepsilon=2.356$ and when we increase the coupling strength to $\varepsilon>2.356$, the attractor H$_2$ attains the complete synchronization manifold, and the symmetry is again preserved in the subsystems. The emerged synchronized attractor is labeled as S (lines with inverted triangles in Fig.~\ref{bifur_al0.01}), and the time series as well as the phase portrait plots of the attractor S are depicted in Figs. \ref{ts_att1}(e) and \ref{ts_att1}(f), respectively, for $\varepsilon=3.0$.
	
	\item Secondly, the QP$_1$ has emerged in the system, if we choose the initial states from another basin of attraction. The two attractors are rotating around the different centers of rotation. We named this attractor as QP$_1$, and this quasiperiodic attractor is stable in the range of the coupling strength  $\varepsilon\in(0.001,0.147]$. The time evolution and the phase space plots of the QP$_1$ attractor are plotted in Figs. \ref{ts_att2}(a) and \ref{ts_att2}(b) for $\varepsilon=0.14$. When we increase the coupling strength further, the QP$_1$ attractor lost its stability through torus bifurcation (marked as TB$_1$ in Fig.~\ref{bifur_al0.01}) and transformed into stable desynchronized limit-cycle oscillations for $\varepsilon>0.147$. The lines with crosses indicate the maxima and minima of the corresponding limit-cycle attractor. Here, both the oscillators have a different center of rotation, which is shown in Figs. \ref{ts_att2}(c) and \ref{ts_att2}(d) for $\varepsilon=0.25$. We named these attractors as IH$_1$ (inhomogeneous oscillations--1). The amplitude of the attractor IH$_1$ is slowly decreasing when one increases the coupling strength, and for $\varepsilon=0.341$, the attractor IH$_1$ is transformed into OD state--2 (attractor OD$_2$) that has emerged via reverse supercritical Hopf bifurcation. The lines with open triangles in Fig.~\ref{bifur_al0.01} represent attractor OD$_2$, and the critical coupling strength of the Hopf bifurcation point is labeled as HB$_2$ in Fig.~\ref{bifur_al0.01}. This steady-state is portrayed in Figs. \ref{ts_att2}(e) and \ref{ts_att2}(f) for $\varepsilon=0.4$. Upon increasing the coupling strength further, the OD state--2 and another unstable fixed point collides with each other and disappeared via saddle-node bifurcation when the coupling strength reaches the value of $\varepsilon=0.653$. The respective bifurcation point is marked as SN$_2$ in Fig.~\ref{bifur_al0.01}.
	
	\item Finally, QP$_2$ attractor exhibits in the range of $\varepsilon\in(0.001,0.22]$. The QP$_2$ attractor also rotates about the different center of rotations. The time series and phase space plots of the QP$_2$ attractor are depicted in Figs. \ref{ts_att3}(a) and \ref{ts_att3}(b), respectively, for $\varepsilon=0.14$ indicating that the amplitude of the QP$_2$ attractor is relatively smaller than QP$_1$. Next, if we increase $\varepsilon$ further, the QP$_2$ attractor is then bifurcated into stable limit-cycle attractor through torus bifurcation (TB$_2$) at $\varepsilon=0.22$ (see Fig.~\ref{bifur_al0.01}). The newly emerged limit-cycles also have two different centers of rotation and oscillate with different amplitudes corroborated from the time evolution and phase portraits, as depicted in Figs. \ref{ts_att3}(c) and \ref{ts_att3}(d), respectively, for $\varepsilon=0.221$. We named this attractor as IH$_2$ (inhomogeneous oscillations--2). These limit-cycles are stable only for a short range of $\varepsilon$ (filled triangles in Fig.~\ref{bifur_al0.01}) and then bifurcated into an OD state--1 (OD$_1$) for $\varepsilon=0.225$ via reverse supercritical Hopf bifurcation, at which the two systems attain different stable, steady-states. For instance, the corresponding states are portrayed in Figs. \ref{ts_att3}(e) and \ref{ts_att3}(f) for $\varepsilon=0.25$. The critical Hopf bifurcation point is marked as HB$_1$ in Fig.~\ref{bifur_al0.01} and the open squares depicted the attractor OD$_1$. After that, the OD$_1$ attractor is lost stability and transformed into a stable limit-cycle attractor (denoted by the solid squares in Fig.~\ref{bifur_al0.01}) for $\varepsilon=0.521$ via supercritical Hopf bifurcation. The newly emerged attractor also have two different origins of rotation, which are named as IH$_4$ (inhomogeneous oscillations--4). The associated time series and phase portraits of the IH$_4$ attractor are depicted in Figs. \ref{ts_att3}(g) and \ref{ts_att3}(h), respectively, for $\varepsilon=0.8$ confirming that the amplitudes are uncorrelated (desynchronized). Finally, only two limit-cycle attractors IH$_4$ (desynchronized) and S (synchronized) are stable and coexisted. 
\end{enumerate}
\begin{figure}
	\includegraphics[width=1.0\columnwidth]{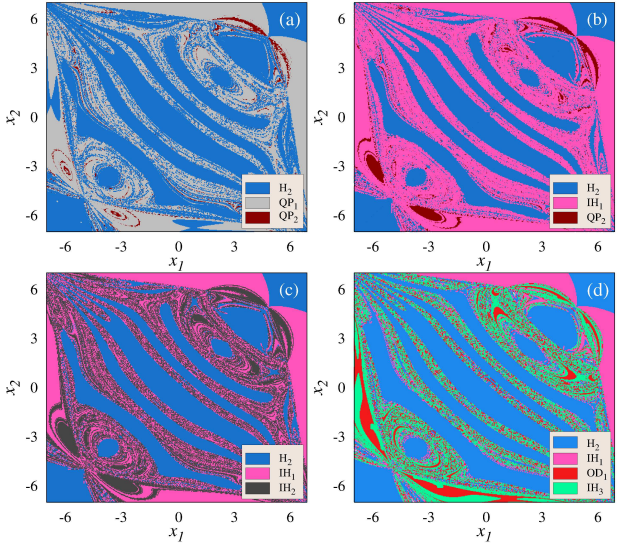}
	\caption{Basins of attraction of the system (\ref{eqn3}) as a function of the state variables $x_1$ and $x_2$ for different values of the coupling strength $\varepsilon$ with $\alpha$=0.01,(a) plotted for $\varepsilon$=0.1, (b) for $\varepsilon$ = 0.18, (c) for $\varepsilon$ = 0.221 and (d) is depicted for $\varepsilon$ = 0.25. The inset shows the riddled nature of the basin in certain ranges of initial states.}
	\label{basin_att_eps1}
\end{figure} 
\begin{figure}
	\includegraphics[width=1.0\columnwidth]{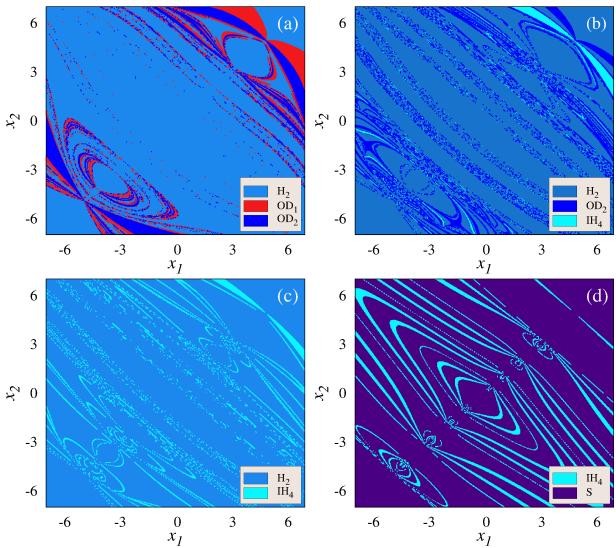}
	\caption{Basins of attraction of the system (\ref{eqn3}) as a function of the state variables $x_1$ and $x_2$ for different values of the coupling strength $\varepsilon$ with $\alpha$=0.01,(a) plotted for $\varepsilon$=0.4, (b) for $\varepsilon$ = 0.6, (c) for $\varepsilon$ = 0.8 and (d) is depicted for $\varepsilon$ = 3.0.}
	\label{basin_att_eps2}
\end{figure} 
\begin{figure*}
	\includegraphics[width=2.0\columnwidth]{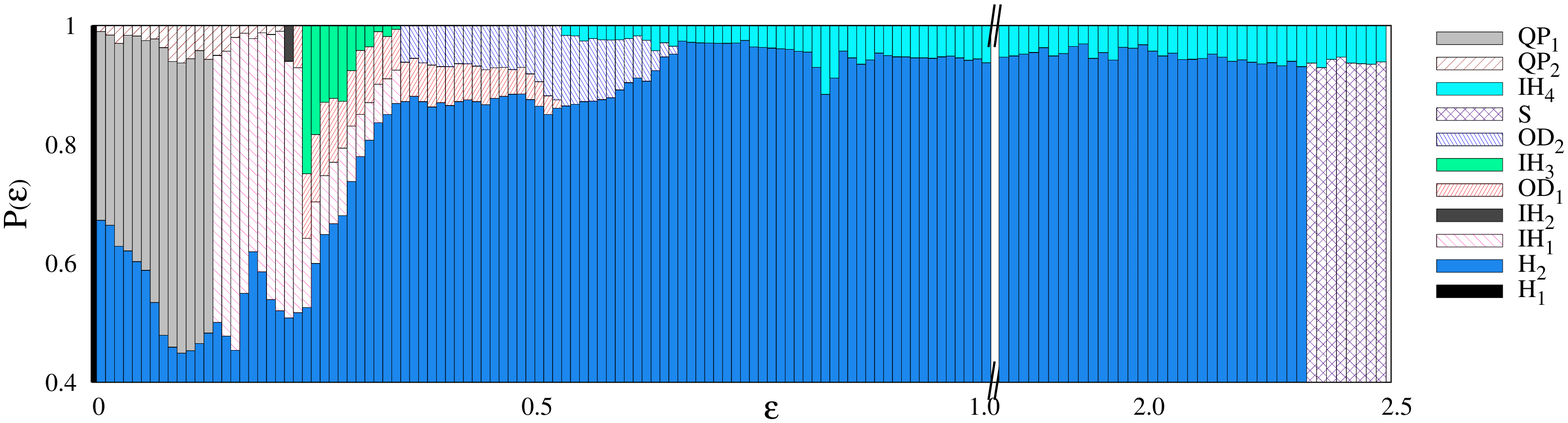}
	\caption{Probability of various attractors from H$_1$ to S including the quasiperioc attractor QP$_1$ and QP$_2$ as a function of the coupling strength and for $\alpha=0.01$. Different colors and patterns indicate the probability of different attractors. }
	\label{prob_att}
\end{figure*} 
Meanwhile, when the coupling strength reaches the value of $\varepsilon=0.245$, an unstable periodic orbit is bifurcated into a stable limit-cycle which is indicated as inhomogeneous oscillations--3 and the corresponding attractors are named as IH$_3$ (solid circles in Fig.~\ref{bifur_al0.01}), in which one of the subsystems oscillate with period-II limit-cycle oscillations, and the other subsystem exhibits period-I limit-cycle oscillations which are evident from Figs.~\ref{ts_att3}(i) and \ref{ts_att3}(j), plotted for $\varepsilon=0.25$. These attractors are stable until the coupling strength reaches the value of $\varepsilon=0.341$. After that, the attractor IH$_3$ is merged with an unstable orbit and disappeared via saddle-node bifurcation (denoted as SN$_1$ in Fig.~\ref{bifur_al0.01}).
\begin{figure*}
\includegraphics[width=2.0\columnwidth]{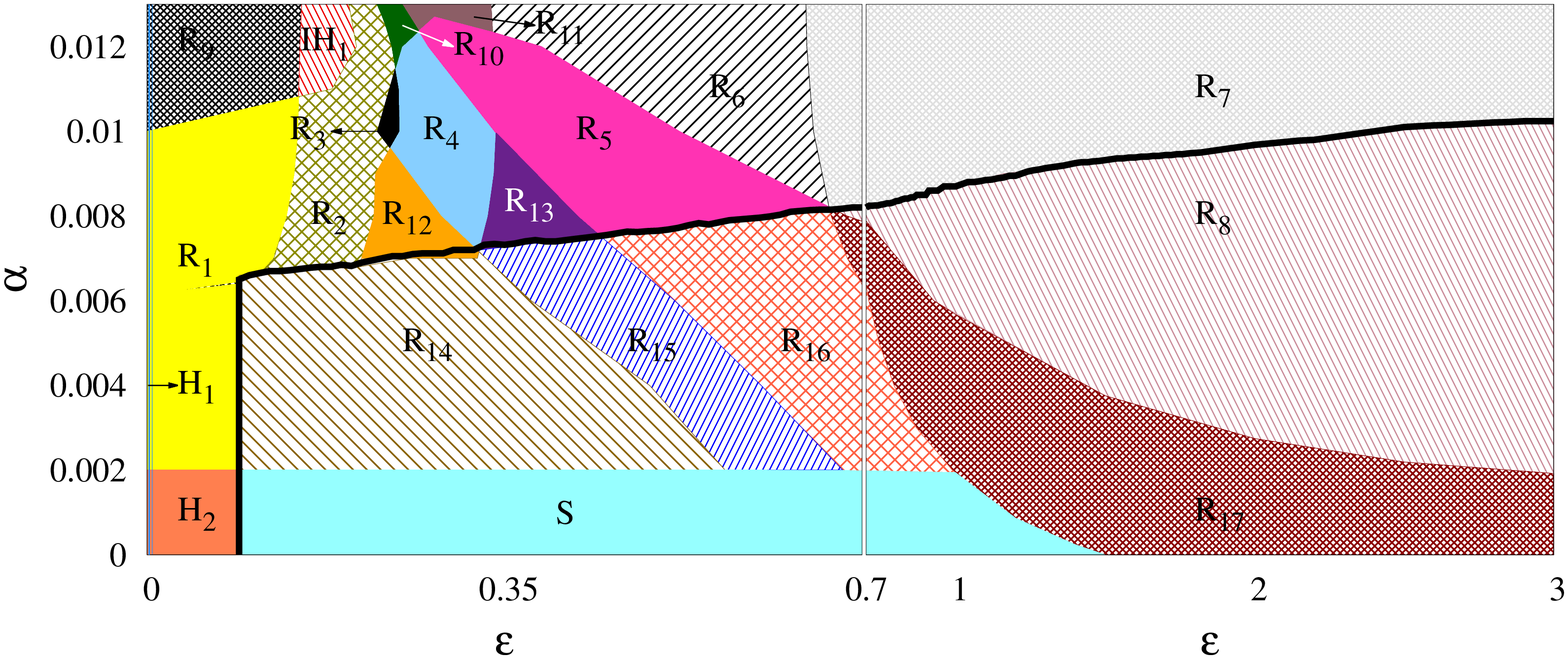}
\caption{Emergence of multistable attractors for different initial states are illustrated in a two-parameter diagram of the coupled R\"ossler oscillator (\ref{eqn3}) as a function of nonisochronicity parameter ($\alpha\in[0,0.013]$) and coupling strength $\varepsilon\in[0,3]$. Distinct colors and patterns represent the combination of different attractors. The combination of attractors in regions R$_1$ to R$_8$ are already illustrated in Sec.~\ref{sec5}. The combination of attractors in other regions from R$_9$ to R$_{17}$ are as follows: R$_9$ indicates the coexistence of attractors H$_2$ and QP$_1$. Similarly, R$_{10}$ = H$_2$+IH$_1$+OD$_1$+IH$_3$, R$_{11}$ = H$_2$+IH$_3$+OD$_2$+IH$_4$, R$_{12}$ = H$_2$+IH$_1$+IH$_3$, R$_{13}$ = H$_2$+IH$_1$+OD$_1$, R$_{14}$ = IH$_1$+S, R$_{15}$ = IH$_1$+OD$_1$+S, R$_{16}$ = OD$_1$+OD$_2$+S, R$_{17}$ = OD$_2$+IH$_4$+S. The think black continuous line indicates the complete synchronization region obtained using MSF formalism (\ref{peq2}). The parameter values are fixed as given in Fig. \ref{rossler_fp}}
\label{2p_att_region}
\end{figure*} 

Thus, the influence of coupling, along with nonisochronicity parameters, is studied in terms of the bifurcation diagram. The emergence of different bifurcations, and coexisting synchronized and desynchronized attractors are discussed. In a nutshell, Table~\ref{tab1} shows the possible attractors that are appeared in the system (\ref{eqn3}) as a function of $\varepsilon\in[0,2.5]$ for $\alpha=0.01$. Next, based on the emergence of different coexisting attractors, the bifurcation diagram is divided into a number of regions, and basins of attraction of each region are described in Sec.~\ref{sec5}.
\section{Basins of attraction in different regions of the bifurcation diagram}
\label{sec5}
Based on the appearance of different bifurcations, Fig.~\ref{bifur_al0.01} is divided into eight regions from R$_1$ to R$_8$, in which different coexisting attractors have emerged for different initial states as a function of the coupling strength. In the following, we discuss the basins of attraction of those regions. It is important to emphasize here that in order to plot the basins of attraction of the coupled system, the initial states of the $x$-variables of the two individual systems are uniformly distributed in the range of $x_{1,2}\in[-7,7]$ and the initial states of the other state variables are fixed at $y_{1,2}=0.5$ and $z_{1,2}=-1$. 
\begin{figure}
	\includegraphics[width=1.0\columnwidth]{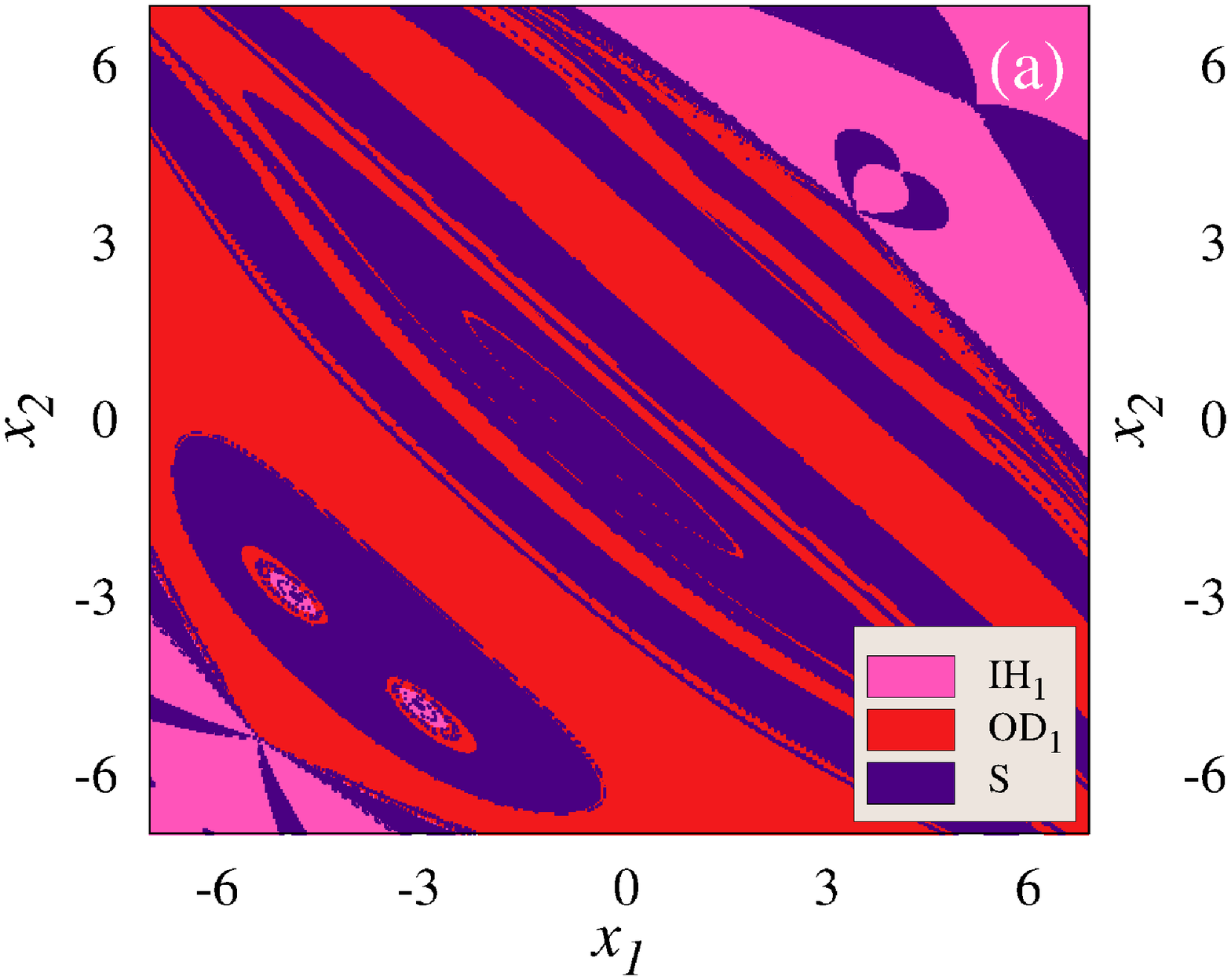}
	\caption{The basins of attraction of the coupled R\"ossler oscillator (\ref{eqn3}) for (a) $\varepsilon=0.5$, and (b) for $\varepsilon=0.6$ with $\alpha=0.006$ shows the coexisting attractors in the regions of R$_{15}$ and R$_{16}$, respectively.}
	\label{basin_states}
\end{figure} 

In region R$_1$, the system exhibits the coexistence of attractors H$_2$, QP$_1$, and QP$_2$. The basins of attraction of the respective attractors are depicted in Fig.~\ref{basin_att_eps1}(a) for $\varepsilon$ = 0.1. In this figure, the blue (dark gray) color region represents the basin of attraction of attractor H$_2$. The light gray color region indicates the basin of attraction of the QP$_1$, and the initial states to which the attractor QP$_2$ has emerged are indicated by the region of maroon (black) color. We can also note that the attractors H$_2$ and QP$_1$ are having an almost equal number of occurrences. The attractor QP$_2$ has occurred for less number of initial states than the other two attractors. Next, in region R$_2$ of the bifurcation diagram, the coupled system exhibits the coexistence of attractors H$_2$, IH$_1$, and QP$_2$, in which IH$_1$ has occurred relatively for a large number of initial states than the other two attractors. The corresponding basins of attraction are plotted in Fig.~\ref{basin_att_eps1}(b) for $\varepsilon=$ 0.18, where at the pink (light gray) color portrays the basin of attraction of IH$_1$ attractor. Next, the attractors H$_2$, IH$_1$ and IH$_2$ coexist in the region R$_3$ for different sets of initial states and the respective basins of attraction is plotted in Fig.~\ref{basin_att_eps1}(c) for the value of coupling strength $\varepsilon$ = 0.221. The dark gray color points illustrate the initial states to which the IH$_2$ has emerged in the system, and IH$_1$ dominates the other two attractors with a high number of occurrences. Further, the system (\ref{eqn3}) shows the coexistence of four different attractors including H$_2$, IH$_1$, OD$_1$, and IH$_3$ in region R$_4$. The basins of attraction of those attractors are depicted in Fig.~\ref{basin_att_eps1}(d) for the value of coupling strength $\varepsilon$ = 0.25 corroborating that H$_2$ has a high number of occurrences. The red (dark gray) color indicates the basin of attraction of OD$_1$, and the light green (gray) color illustrates the basin of attraction of the attractor IH$_3$.%

Moreover, in region R$_5$, the coupled system exhibits the coexistence of two OD states (OD$_1$ and OD$_2$) with attractor H$_2$. For instance, the corresponding basins of attraction is plotted in Fig.~\ref{basin_att_eps2}(a) for $\varepsilon=0.4$ in which the dark blue (black) color region describes the basin of attraction of the attractor OD$_2$. Next, the basins of attraction in the region R$_6$ are plotted in Fig.~\ref{basin_att_eps2}(b) for $\varepsilon=0.6$ in that the system shows the coexistence of the attractors H$_2$, OD$_2$, and IH$_4$. The aqua (light gray) color indicates the basin of attraction of IH$_4$. Furthermore, the basins of attraction of region R$_7$ is depicted in Fig.~\ref{basin_att_eps2}(c) for $\varepsilon$ = 0.8 at which only the attractors A$1$ and IH$_4$ are co-occurred. In all these Figs.~\ref{basin_att_eps2}(a),\ref{basin_att_eps2}(b) and \ref{basin_att_eps2}(c) the attractor H$_2$ dominates the other attractors with a high number of occurrences. Finally, in region R$_8$, the newly emerged synchronized attractor S has coexisted with IH$_4$, and dominates the basin of attraction with a high number of occurrences, which is plotted for $\varepsilon$ = 3.0 in Fig.~\ref{basin_att_eps2}(d) .

Therefore, from Figs.~\ref{basin_att_eps1} and \ref{basin_att_eps2}, one can observe that each attractor dominates the other attractors with a high number of occurrences for large values of initial states. Therefore it is also of interest to investigate the probability of different attractors with respect to the coupling parameter $\varepsilon$\cite{menck2013,schultz2017,brzeskit2019}. Figure ~\ref{prob_att} shows the probability of various attractors for 500 uniformly distributed initial states of $x_{1,2}\in$ [-7, 7] for $\alpha=0.01$ and as a function of $\varepsilon$. From Fig.~\ref{prob_att}, one can easily uncover that the attractor H$_1$ predominantly occurs in the probability diagram with a high number of occurrences and shares the basin of attraction with other coexisting attractors for $\varepsilon<2.356$. After that, the S attractor dominates the basin to our calculated range of $\varepsilon=3.0$ and shares the parameter space with IH$_4$.    

To clarify the coexistence of different attractors, we have plotted a two-phase diagram in the parameter space of $\varepsilon\in[0, 3]$ and $\alpha\in[0, 0.013]$ in Fig.~\ref{2p_att_region}, which depicts the multistability of attractors for different values of $\varepsilon$ and $\alpha$. Each colored and patterned regions indicate the coexistence of different attractors. One can also notice that, apart from the regions R$_1$ to R$_8$ as discussed earlier, there exist other combinations of coexisting attractors in regions from R$_9$ to R$_{17}$ which are also indicated in Fig.~\ref{2p_att_region}. For an illustration, in region R$_{15}$, the attractors IH$_1$, OD$_1$ and S coexist and the respective basins of attraction is plotted in Fig.~\ref{basin_states}(a) for $\varepsilon=0.5$ and $\alpha=0.006$. Similarly, the basins of attraction of the region R$_{16}$ is plotted in Fig.~\ref{basin_states}(b) for $\varepsilon=0.6$ and $\alpha=0.006$, in which OD$_1$ and OD$_2$ are coexisting along with the synchronized attractor S. 

The region of the stable synchronized state for the coupled system (\ref{eqn3}) is identified using the well-known MSF formalism. The complete synchronization manifold is defined by $x_1=x_2=x$, $y_1=y_2=y$ and $z_1=z_2=z$, where $x,y,z$ are the solutions of the uncoupled system (\ref{eqn1}). Let, $(\eta_{j},\xi_{j},\zeta_{j})$ are the deviations of $(x_{j}, y_j,z_j)$, $j=1,2$ from the synchronized solution $(x,y,z)$ and we have 
\begin{eqnarray}
\dot{\eta}_{1,2}&=&-\omega_0(\xi_{1,2}-\alpha (2 x y \eta_{1,2}+(x^2 +3y^2)\xi_{1,2}))-\zeta_{1,2}+\nonumber\\ & &\epsilon (\eta_{2,1}-\eta_{1,2})  \nonumber\\
\dot{\xi}_{1,2}&=&\omega_0(\eta_{1,2}-\alpha (2y x\xi_{1,2}+(3x^2+y^2)\eta_{1,2}))+a \xi_{1,2}\nonumber\\
\dot{\zeta}_{1,2}&=&x\zeta_{1,2}+z\eta_{1,2}-c \zeta_{1,2}.
\label{peq1}
\end{eqnarray}
From Eq.~(\ref{peq1}) we get, 
\begin{eqnarray}
\dot{\eta}&=&-\omega_0(\xi-\alpha (2 x y \eta+(x^2 +3y^2)\xi))-\zeta-2\epsilon \eta \nonumber\\
\dot{\xi}&=&\omega_0(\eta-\alpha (2y x\xi+(3x^2+y^2)\eta))+a \xi\nonumber\\
\dot{\zeta}&=&x\zeta+z\eta-c \zeta,
\label{peq2}
\end{eqnarray}
where $(\eta,\xi,\zeta)=(\eta_{1}-\eta_{2},\xi_{1}-\xi_2,\zeta_{1}-\zeta_2)$ are the transverse small perturbations on the synchronization manifolds.  The synchronized state is stable when all the transverse perturbations $(\eta,\xi,\zeta)$ tend to zero when $t\rightarrow \infty$. This will happen when all the transverse Lyapunov exponents are negative. Using the equation (\ref{peq2}) we have identified the region of stable complete synchronization state in ($\varepsilon, \alpha$) parameter space. The thick black continuous line in Fig.~\ref{2p_att_region} represents the region of stable complete synchronization.
\section{Conclusions}
\label{sec7} 
In summary, we have systematically studied the influence of nonisochronicity on the single and mutually coupled R\"ossler oscillator. First, we have examined the fixed point analysis of a single R\"ossler oscillator with and without nonisochronicity term. We showed that adding the nonisochronicity term can increase the number of fixed points, which induces multistability of coexisting limit-cycle and quasiperiodic attractors. Next, we have extended our studies to a system of two mutually coupled R\"ossler oscillators and investigated the influence of the coupling parameter along with the nonisochronicity parameter. We showed that the permutational symmetry of the coupled system is broken when one increases the coupling strength, persuades multistability of desynchronized attractors, which are emerged via different bifurcations. The broken symmetry is preserved again in the system when the coupling strength reaches higher values, engenders the emergence of a synchronized attractor. In addition to the synchronized and desynchronized attractors, we have reported the emergence of two different OD states. The results are corroborated using the bifurcation diagrams, and basins of attraction. The region of complete synchronization is identified using the MSF formalism. 

The results presented in the paper are robust against system parameters, and similar results can be observed even if the systems exhibit chaotic dynamics. Also, we note that the results are obtained only in the two coupled systems. However, it is also of interest to understand the interplay of nonisochronous nature of the system and coupling topology in different network architectures to induce rich collective dynamical behaviors and to identify the network topologies that enhance or support the effects of nonisochronicity are still open problems, and we are processing our research in this direction.

\begin{acknowledgements}
C. Ramya acknowledges SASTRA for financial support in the form of Teaching Assistantship. The work of R. S. is supported by the SERB-DST Fast Track scheme for young scientists under Grant No. YSS/2015/001645. The work of V. K. C. and R. G. are sponsored by the SERB-DST-CRG Grant No. CRG/2020/004353 and CSIR EMR Grant No. 03(1444)/18/EMR-II. All the authors thank DST, New Delhi for computational facilities under the DST-FIST programme (SR/FST/PS-1/2020/135) to the Department of Physics. 
\end{acknowledgements}
\vskip 10pt
\noindent \textbf{DATA AVAILABILITY}\\
The data that support the findings of this study are available from the corresponding author upon reasonable request.\\	

\end{document}